# Wind, wave and current interactions appear key for quantifying cross-shelf transport and carbon export; new knowledge and the potential of SKIM to enable monitoring

Jamie D. Shutler[1]*, Thomas Holding[1], Clement Ubelmann[2], Lucile Gaultier[3], Fabrice Collard[3], Fabrice Ardhuin[4], Bertrand Chapron[4], Marie-Helene Rio[5], Craig Donlon[6]

* Corresponding author

[1] University of Exeter, Centre for Geography and Environmental Sciences (CGES), Penryn, UK.

[2] Collecte Localisation Satellites (CLS), Brest, France.

[3] OceanDataLab (ODL), Brest, France.

[4] Ifremer, Laboratoire d'Oceanographie Physique et Spatiale (LOPS), IUEM, Brest, France.

[5] European Space Agency (ESA), Frascati, Italy.

[6] European Space Agency (ESA), Noordwijk, The Netherlands.

**Abstract**

The highly heterogeneous and biologically active continental shelf-seas are important components of the oceanic carbon sink. Carbon rich water from shelf-seas is exported at depth to the open ocean, a process known as the continental shelf pump, with open-ocean surface water moving (transported) onto the shelf driving the export at depth. Existing methods to study shelf-wide exchange focus on the wind or geostrophic currents, often ignoring their combined effect, spatial heterogeniety or any other ageostrophic components. Here we investigate the influence that wind, wave and current interactions can have on surface transport and carbon export across continental shelves. Using a 21 year global re-analysis dataset we confirm that geostrophic and wind driven Ekman processes are important for the transport of water onto shelf seas; but the dominance of each is location and season dependent.  A global wave model re-analysis shows that one type of ageostrophic flow, Stokes drift due to waves, can also be significant. Furthermore, a regional case study using two submesocale model simulations identifies that up to 100% of the cross-shelf surface flow in European seas can be due to ageostrophic components.  Using these results and grouping shelf-seas based on their observed carbon accumulation rates shows that differences in rates are consistent with imbalances between the processes driving atmosphere-ocean exchange at the surface and those driving carbon export at depth. Therefore expected future changes in wind and wave climate support the need to monitor cross-shelf transport and the size of the continental shelf-sea carbon pump.  The results presented show that a new satellite concept, the Sea Surface Kinematics Multiscale monitoring satellite mission (SKIM), will be capable of providing measurements of the total cross-shelf current, which are now needed to parameterise models and enable routine monitoring of the global continental shelf-sea carbon pump.

**Social media abstract**

Wind, wave & currents appear key for quantifying cross-shelf transport & carbon export;SKIM can help

**1.0 Introduction**

Ocean surface wind, currents and waves and their interactions play a major role in governing the exchange of heat, energy and carbon between the atmosphere and the ocean, and their onward transport. Ocean currents within the global oceans are a combination of geostrophic currents ($C_G$) due to changes in pressure and the Coriolis effect, Ekman currents ($C_E$) due to wind stress on the surface water, tides and other ageostrophic components including Stokes drift (the radial motion due to surface waves) and eddies (rotating volumes of water). The Ekman layer depth is the penetration depth of wind-induced turbulent mixing and this sets up the mixed layer depth, the vertical layer of water at the surface where all constituents including salinity, temperature, oxygen and nutrients are uniformly distributed. The currents across multiple depths within this mixed layer are the dominant drivers of the net transport of water across the shelf-edge and into shelf seas, which in turn, through the need for mass balance, drives the reciprocal and opposing flow back to the open ocean. In shelf seas this opposing flow will tend to occur at depth, and provides the conduit for carbon export from the shelf-sea to the deep ocean, a process referred to as the continental shelf-sea carbon pump.

The transport of water onto shelf seas in the surface mixed layer is often calculated using either geostrophy (the balance between pressure gradient forces and Coriolis forces) or wind speed measurements, despite the knowledge that wind, wave and current interactions can play a major role in driving transport. This simplification is invariably due to the lack of observations available to accurately characterise all processes that govern transport across the shelf. As a result approaches for calculating this surface transport tend to assume that wind driven Ekman currents dominate at all times and locations (e.g. Painter et al., 2016), or that geostrophic currents dominate (e.g. Yuan et al., 2017), whilst some identify the importance of a combination of Ekman, geostrophic and ageostrophic components including the role of Stokes drift (Feewings et al., 2008; Woodson, 2013) or eddies (Waite et al., 2016). All of these studies are either based on sparse *in situ* data collection and measurements, or single satellite altimeter returns, so the assessments are very coarse in space and/or time. Therefore they provide a good analysis for the region or time period studied, but they are either i) unable to identify the influence and significance of interactions between different geostrophic and ageostrophic current components (e.g. Painter et al., 2016; Yuan et al., 2017) or ii) unable to identify how this alters in space and time (Feewings et al., 2008; Woodson, 2013; Waite et al., 2016); all of which are important when quantifying the net transport of water onto the shelf.

Faster exchange of shelf-sea carbon with the ocean interior combined with biological activity within shelf-seas may have helped slow down the rate of increase in surface water partial pressure of $CO_2$ ($pCO_2$) in many shelf regions (Laruelle et al., 2018). Two mechanisms have been proposed to explain how the continental shelf $CO_2$ sink has evolved (Laruelle et al., 2018). The first is based on the atmosphere-ocean exchange at the water surface and its later export to the deep ocean at depth. Imbalances between these two exchange processes could modulate the carbon accumulation (ocean acidification) within shelf seas (Cai, 2011; Bauer et al., 2013). The second is the evolution of the biological pump due to anthropogenic nutrient inputs resulting in a change from net heterotrophy (organism dependent on complex organic substances) to net autotrophy (organisms that can synthesise their own food) (Bourgeois et al., 2016).

This study is interested in characterising the controls of surface transport onto continental shelves and placing this into context for carbon and ecosystem health assessments. A recently developed globally resolved ocean current dataset that includes geostrophic and wind driven Ekman currents at two depths (Rio et al., 2014) provides an opportunity to assess the validity of using a purely Ekman or purely geostrophic focussed analysis. This combined with a wave model re-analysis of Stokes drift data allows a global assessment of the importance of Ekman, geostrophic and one ageostrophic current component for driving cross-shelf transport. All of these results are then placed into context for fourteen continental shelf-seas that are exhibiting differing rates of change in surface water $pCO_2$. Collectively this work supports the hypothesis that imbalances between atmosphere-ocean exchange and carbon export at depth are controlling the change in shelf-sea $CO_2$ sinks and their acidification rates.

The Sea Surface Kinematics Multiscale monitoring satellite mission (SKIM, Ardhuin et al., 2018) is an Earth Explorer-9 (EE9) candidate mission and if launched will provide the first direct measurements of total surface ocean velocity and wave spectrum. This means that SKIM will have the capability to provide direct measurements of cross-shelf surface water velocities. A case study focussing on the European shelf region is used to identify the importance of ageostrophic current components for controlling cross-shelf transport of water in European seas and the potential for SKIM to observe them. To date, no method for observing global cross-shelf ageostrophic flow exists. This study suggests that such a capability is necessary if we are to monitor the strength of the continental shelf sea pump and the impact of a changing climate on this pump towards motivating societal shifts needed for meeting carbon emission targets.

Section 2 describes the cross-shelf transport and ocean current analysis methods, and the results are presented in section 3. The implications for the shelf sea carbon export and the potential for routine monitoring of the shelf sea continental pump is presented in section 4. Section 5 provides the conclusions from this work.

**2.0 Methods**

Unless otherwise stated all analyses were performed on global datasets.

**2.1 Determining the shelf boundary and current vectors normal to the shelf edge**

Shelf boundaries were identified as the 500 m depth contour calculated from GEBCO bathymetry data (Weatherall et al., 2015) and resampled to match the 0.25° spatial resolution of the ocean current datasets (Figure 1a). A filter was used to remove short paths (paths <150 points) due to islands and inland seas from the shelf-edge boundary. To calculate the onto-shelf (ocean) current vectors each path in the shelf boundary was divided into *n* equal-distance segments approximated using straight lines (figures 1b and 1c). The onto-shelf direction vector was then determined using the normal line to each straight-line segment in the direction that first bisects a deeper contour line (600 m) before then bisecting the shallower contour line (500 m). Each point along the shelf boundary was then assigned an onto-shelf direction vector corresponding to the nearest straight line segment (figures 1b and 1c). Larger values of *n* lead to improved capability to follow complex shelf contours (e.g. compare figures 1b to 1c), but with increased computation time. *n* was chosen as a function of the number of points ($N_p$) in each contour path (p) that make up the shelf boundary, using n = $(0.05N)+1_p$. Figure S1 shows all resulting shelf boundaries. This approach provides the coordinates and onto-shelf direction for each point along the shelf boundary line. Next the locations where the shelf boundary intersects grid lines on a 0.25° grid are determined, allowing current velocities to be extracted from the relevant ocean current data point along each shelf boundary line (figure 1d). To assess the sensitivity of this approach to the choice of shelf boundary depth the cross-shelf currents were also calculated using shelf boundary depths of 300 to 700 metres (in 50 m steps). In each case the deep contour depth (see section 3.1) was selected to be 100 m deeper than the shelf boundary depth. The sensitivity was determined by calculating the standard deviation of the cross-shelf current due to differing shelf boundary depths.

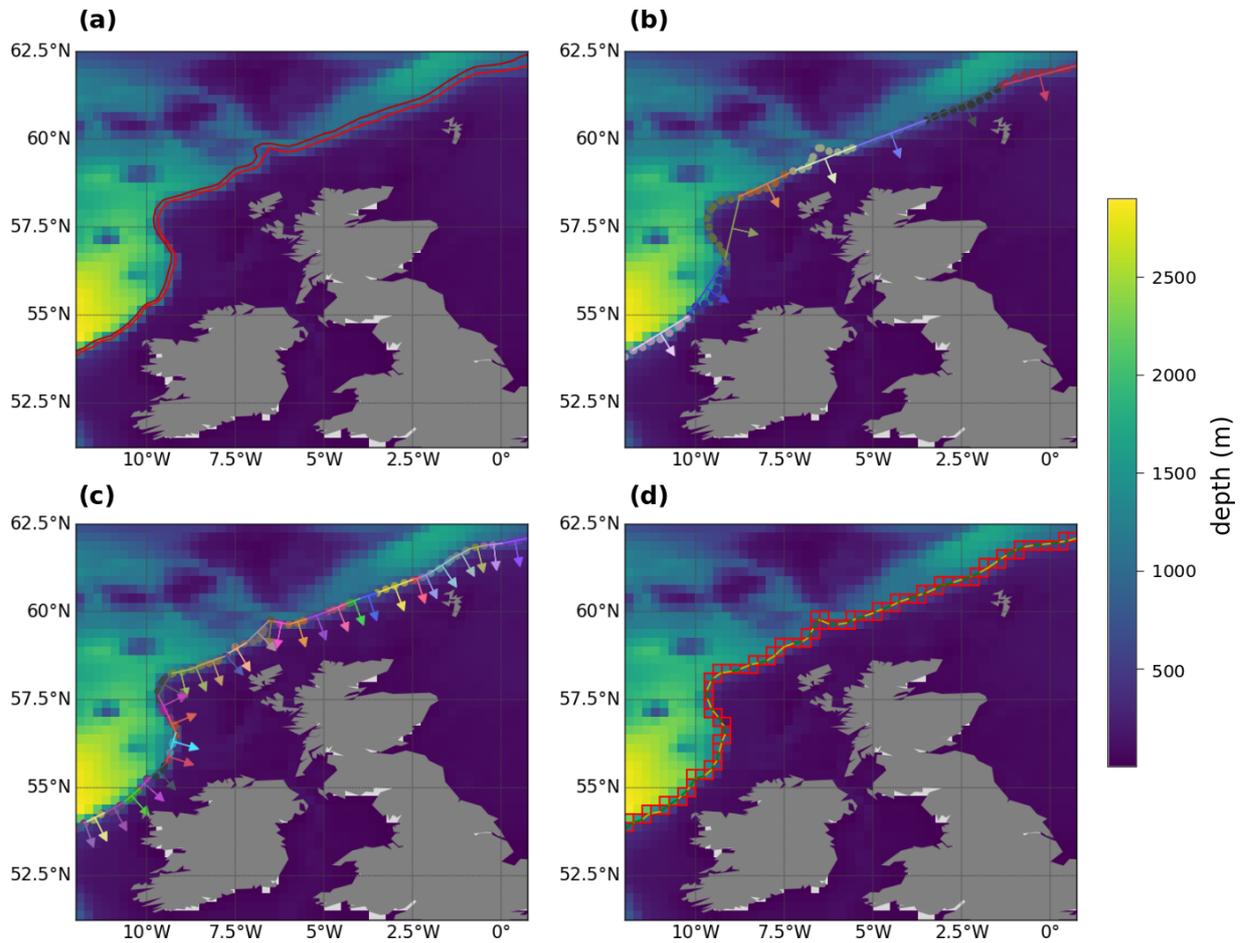

**Figure 1:** Step by step visualisation of the method for extracting shelf boundary information. **(a)** Two contour lines are extracted to represent the shelf boundary: the shallow (500 m, bright red) and deep (600 m, dark red) contour. **(b)** and **(c)** The shelf boundary is approximated by *n* of straight lines (here *n*=8 and 64), as indicated by the coloured straight lines with the calculated onto-shelf direction shown by arrows. Each point along the boundary (shallow contour line) is assigned an onto-shelf direction vector of the associated straight line (this is indicated by paired colours between straight lines and contour segment). **(d)** Identifying each data point or grid cell (red boxes) along the shelf boundary.

**2.2 Indicator of Ekman versus geostrophic dominance of across-shelf exchange**

The Ekman current ($C_E$) both rotates and decreases in magnitude with depth, so $C_E$ peaks in magnitude at the surface. The net transport across the mixed layer due to Ekman processes will be at ~90° to the direction of the wind, or ~45° to the direction of the upper-most component of $C_E$. The magnitude of the surface component of $C_E$ offset by 45° and normal to the shelf edge, $|\mathbf{n}(C_E + 45°)|$, provides an estimate of the upper range of the current strength crossing the shelf-edge within the mixed layer. The geostrophic current ($C_G$) is comprised of barotropic and baroclinic components. The former are depth independent and the latter are density dependent and so can vary with temperature and salinity, and thus depth. However, the density within the mixed layer will be approximately uniform, so a surface observed $C_G$ should be valid for all depths within the mixed layer. Therefore the ratio of $|\mathbf{n}(C_E + 45°)|$, to $|\mathbf{n}(C_G)|$ is used here to indicate the geographic locations and temporal periods where the geostrophic or Ekman current components dominate the cross-shelf transport within the mixed layer. Monthly mean geostrophic and Ekman current data (1993-2016) were derived from the GlobCurrent re-analysis product (Chapron, 2015; Rio et al., 2014; v3, global, 0.25° × 0.25°, 3 hourly). The method in section 2.1 was used to identify the surface currents components and determine $\mathbf{n}(C_E + 45°)$ and $n(C_G)$, according to:

$$\mathbf{n}(C_E + 45°) = \mathbf{D} \cdot \mathbf{E}_{+45} \qquad (1)$$

$$n(C_G) = \mathbf{D} \cdot \mathbf{G} \qquad (2)$$

where $\mathbf{G}$ and $\mathbf{E}$ are vectors containing the North and East components of surface current for geostrophic and Ekman currents, respectively; $\mathbf{D}$ is a unit vector describing the onto-shelf direction. The relative dominance, d, across the shelf edge was then calculated as the ratio:

$$d = \frac{|n(C_G)|}{|n(C_{E+45})| + |n(C_G)|} \qquad (3)$$

Finally, the net surface flow across the shelf boundary is calculated by multiplying the current at each grid point by the distance of the shelf boundary passing through that grid cell, and then summing over all grid points and shelf boundary sections.

**2.3 Indicator of Stokes drift influence on across-shelf transport**

The degree of turbulent flow at the surface (Reynolds stress term) is generally dominated by wind fluctuations and wind stress but can also encompass a wave (orbital motion)-induced stress. The gradient of this wave-induced stress leads to a surface drift, the Stokes drift. Its strength decreases exponentially with depth, but can still be highly correlated with offshore transport (Woodson, 2013). The depth dependency means that the maximum influence will occur within shallow mixed layers that exist during low wind (weak Ekman) conditions. Following Woodson (2013) such conditions are defined as low along-shelf wind stress ($|\tau^{Wy}| < 0.03$ N m$^{-2}$) coincident with large significant wave heights ($H_s > 2$ m), where wind stress is defined as $\tau = \rho_a\, C_d\, (U_{10})^2$, $\rho_a$ is the density of air, $C_d$ is the drag coefficient and $U_{10}$ is the 10 m wind speed. Regions and temporal periods were Stokes drift could be a significant controller of cross-shelf transport were identified using these conditions and a global coverage wave model re-analysis dataset (Rascle and Ardhuin, 2013; WAVEWATCH III development group, 2016). The mathematical method described in section 2.1 was used to determine the shelf boundary and normal vectors. The proportion of total surface current across the shelf edge due to the Stokes drift current, $C_S$, was then calculated as:

$$P_{Stoke} = \frac{|n(C_S)|}{|n(C_{E+45})| + |n(C_G)| + |n(C_S)|} \tag{4}$$

**2.4 Case study: cross-shelf currents in European continental shelf seas**

The case study focuses on the shelf-edge within the North East Atlantic between 42 to 62° N.

**2.4.1 Case study data**

The North Atlantic configuration (NATL60) of the Nucleus for European Modelling of the Ocean (NEMO, v3.6; Madec, 2016) was designed to capture and simulate ocean sub-mesoscale features and flows. The configuration used here (CJM165) includes the coupling of an ice model, atmospheric forcing and boundary conditions (MEOM Group, 2018). This combined simulation (NATL60 CJM165, hourly, 1/60° at the equator, 300 vertical levels) for one calendar year (October 2011 to September 2012) was used as the reference conditions. The geostrophic and Ekman current component data resulting from three altimeters flying over these reference conditions were determined using the methods of Pujol et al., (2016) and Rio et al., (2014) respectively. The total surface current velocities resulting from SKIM flying over the reference conditions were simulated using the SKIMulator framework (version 1.31, with Gaussian surface current noise with a mean of 0.02 ms$^{-1}$; software and documentation are available at https://github.com/oceandatalab/skimulator).

**2.4.2 How important is the ageostrophic component in European shelf seas?**

The mathematical methods described in section 2.1 and 2.2 were used to determine the shelf boundary and normal vectors enabling the magnitude of the simulated reference total current (the simulated truth NATL60 CJM165 dataset) and respective geostrophic and Ekman components to be calculated. The signed percentage difference between the reference and the geostrophic and Ekman components was used to identify the strength of the residual component. The residual, the component of the current that is unaccounted for by geostrophy and Ekman, is then assigned as the ageostrophic component. This first approximation of the ageostrophic component assumes that all current components sum linearly.

**2.4.3 The potential of SKIM to resolve the total cross-shelf currents**

The methods of 2.4.2 were repeated using the simulated SKIM (total current) dataset as the reference. The signed percentage difference between the SKIM reference and the geostrophic and Ekman components (from section 2.4.2) was used to determine the ability of SKIM to capture the ageostrophic component.

## 3.0 Results

### 3.1 Verification of the method

Table 1 shows the shelf exchange controls of different shelf-seas as identified from studies in the literature, in comparison to the calculated results using the method presented here. Where possible the temporal and spatial constraints have been matched. Overall the method presented here produces results that are consistent with the results from these *in situ* studies. For example, it is reassuring to see that the Californian coast results (upwelling and non-upwelling seasons) correctly identifies that Ekman dominates and Stokes is insignificant during the upwelling season, whereas Stokes becomes significant during the non-upwelling season. The notable exception is the work of Painter et al., (2016) who assumed that the cross-shelf transport was Ekman dominated, but no measurements were taken to verify this. The results in Table 1 suggest that the omission of both Stokes and geostrophic current components in their analysis is one explanation for the discrepancy between their wind speed estimated onto shelf transport versus their measured off-shelf transport. The total cross-shelf surface current values are included in Table 1 for interest only; the calculated total cross-shelf currents in all cases where comparison data are available appear consistent with the values from the *in situ* studies. The results investigating the sensitivity of the cross-shelf currents to the depth definition of the shelf edge are presented in table S1 and show close agreement across all shelf-edge definitions in all cases.

### 3.2 Cross-shelf exchange in global shelf seas

Figure 2 identifies where geostrophic currents within the mixed layer and normal to the shelf-edge dominate exchange over Ekman currents (i.e. the results from applying equation 3) for the northern hemisphere summer (Figure 2a) and winter (Figure 2b) calculated across years 1993 to 2014. Figure 3 shows the individual contributions to exchange within the mixed layer and normal to the shelf edge for a selection of shelf edges. Some regions are clearly dominated by geostrophic cross-shelf exchange (e.g. The North Sea in Figure 2 and Figure 3a), whereas the cross-shelf exchange in many regions is a mixture of both geostrophic and Ekman processes (e.g. English Channel and Tasmanian Shelf in Figure 2 and Figures 3a and 3b). Some regions also exhibit distinct changes in hydrodynamic regimes along the shelf (e.g. the Mid Atlantic Bight in Figure 2 and Figure 3c). Figure 4 identifies the regions and periods when ageostrophic Stokes drift could become important for cross-shelf exchange (the results from applying Equation 6). Generally Stokes drift contributes a smaller component to surface cross-shelf exchange than geostrophy or

Ekman processes, but its contribution can still equal (e.g. Californian Coast non-upwelling period, Table 1) or exceed that of Ekman (year round on parts of the European shelf, Table 1, Figure 3a and 4a and 4b).

**Table 1** Verification of cross-shelf transport results against the literature. Where possible the temporal period between the calculated components are consistent with the period studied in the literature.

| *In situ* Region, Reference (indication of measurement depth if known) | *In situ* identified controls or assumed control (* dominance if determined) | *In situ* study period (*In situ* cross-shelf current mean or range, ms$^{-1}$) | This study: Calculated cross-shelf current (ms$^{-1}$) | This study: identified components (proportion within the mixed layer) |
|---|---|---|---|---|
| Scottish Hebrides, Painter et al., (2016) (surface to 2100 m) | Ekman (assumed *) | September 2014 (0.15 – 0.36) | 0.19 (+/- 0.05) | Geostrophic (0.21 +/- 0.15) Ekman (0.16 +/- 0.03) Stokes (0.62 +/- 0.15) |
| South Atlantic Bight, Yuan et al., (2017), (surface, 0 to ~5 m) | Geostrophic, Ekman | January to March, 2002 to 2014 | -0.53 (+/- 0.55) | Geostrophic (0.90 +/- 0.14) Ekman (0.10 +/- 0.14) Stokes (0.01 +/- 0.04) |
| South Atlantic Bight, Yuan et al., (2017), (surface, 0 to ~5 m) | Geostrophic, Ekman | July to September, 2002 to 2014 | -0.58 (+/- 0.61) | Geostrophic (0.92 +/- 0.11) Ekman (0.08 +/- 0.11) Stokes (<0.01 +/- 0.01) |
| Mid Atlantic Bight, Fewings et al., (2008), (surface, 0 to 12 m) | Ekman, Stokes (Geostrophic ignored due to close proximity to shoreline) | October-March, 2001 to 2007 (0.01) | 0.07 (+/- 0.09) | Geostrophic (0.59 +/- 0.24) Ekman (0.26 +/- 0.21) Stokes (0.15 +/- 0.20) |
| Mid Atlantic Bight, Fewings et al., (2008), (surface, 0 to 12 m) | Ekman, Stokes (Geostrophic ignored due to close proximity to shoreline) | April-September, 2001 to 2007 (<0.06) | 0.04 (+/- 0.08) | Geostrophic (0.66 +/- 0.23) Ekman (0.30 +/- 0.23) Stokes (0.05 +/- 0.13) |
| California coast, Woodson, (2013), (surface, 0 to 21 m) | Ekman*, Stokes | April-September, upwelling season, 2004 to 2009 | 0.13 (+/- 0.06) | Geostrophic (0.20 +/- 0.14) Ekman (0.74 +/- 0.20) Stokes (0.06 +/- 0.14) |
| California coast, Woodson, (2013), (surface, 0 to 21 m) | Ekman, Stokes* | October-March, non-upwelling season, 2004 to 2009 | 0.08 (+/- 0.07) | Geostrophic (0.25 +/- 0.16) Ekman (0.39 +/- 0.21) Stokes (0.36 +/- 0.19) |
| Eastern Indian Ocean, Western Australia, Waite et al., (2016), (surface to 400 m) | Eddies (only ageostrophic components were only studied). | May 2006 (autumn) | 0.11 (+/- 0.21) | Geostrophic (0.80 +/- 0.18) Ekman (0.09 +/- 0.08) Stokes (0.11 +/- 0.11) |
| East China Sea, Wei, (2013), (surface to 1200 m) | Geostrophic, Ekman | April 1987 to January 2010 | -0.14 (+/- 0.26) | Geostrophic (0.78 +/- 0.22) Ekman (0.22 +/- 0.22) Stokes (<0.01 +/- 0.03) (calculated for January 1993 to January 2010) |

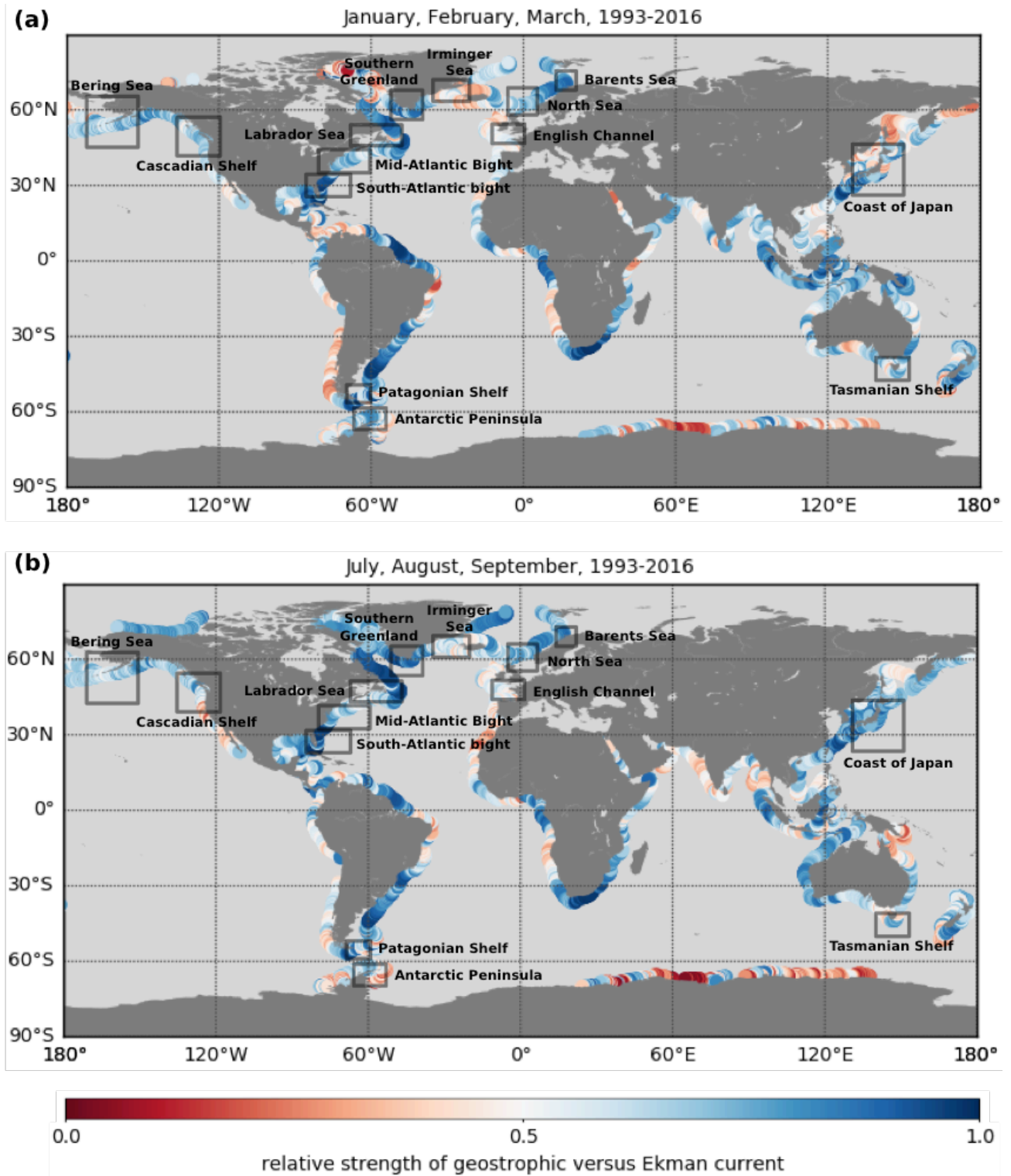

**Figure 2** Relative strength indicator results for the geostrophic versus Ekman transport across all continental shelf sea boundaries for 1993 to 2016, with shelf-seas of interest labelled during a) northern hemisphere winter (January, February and March) and b) during northern hemisphere summer (July, August and September). A value of 0.0 indicates transport dominated by Ekman processes, whereas a value of 1.0 indicates transport dominated by geostrophic components. The boxes are the shelf-seas studied by Laruelle et al., (2018).

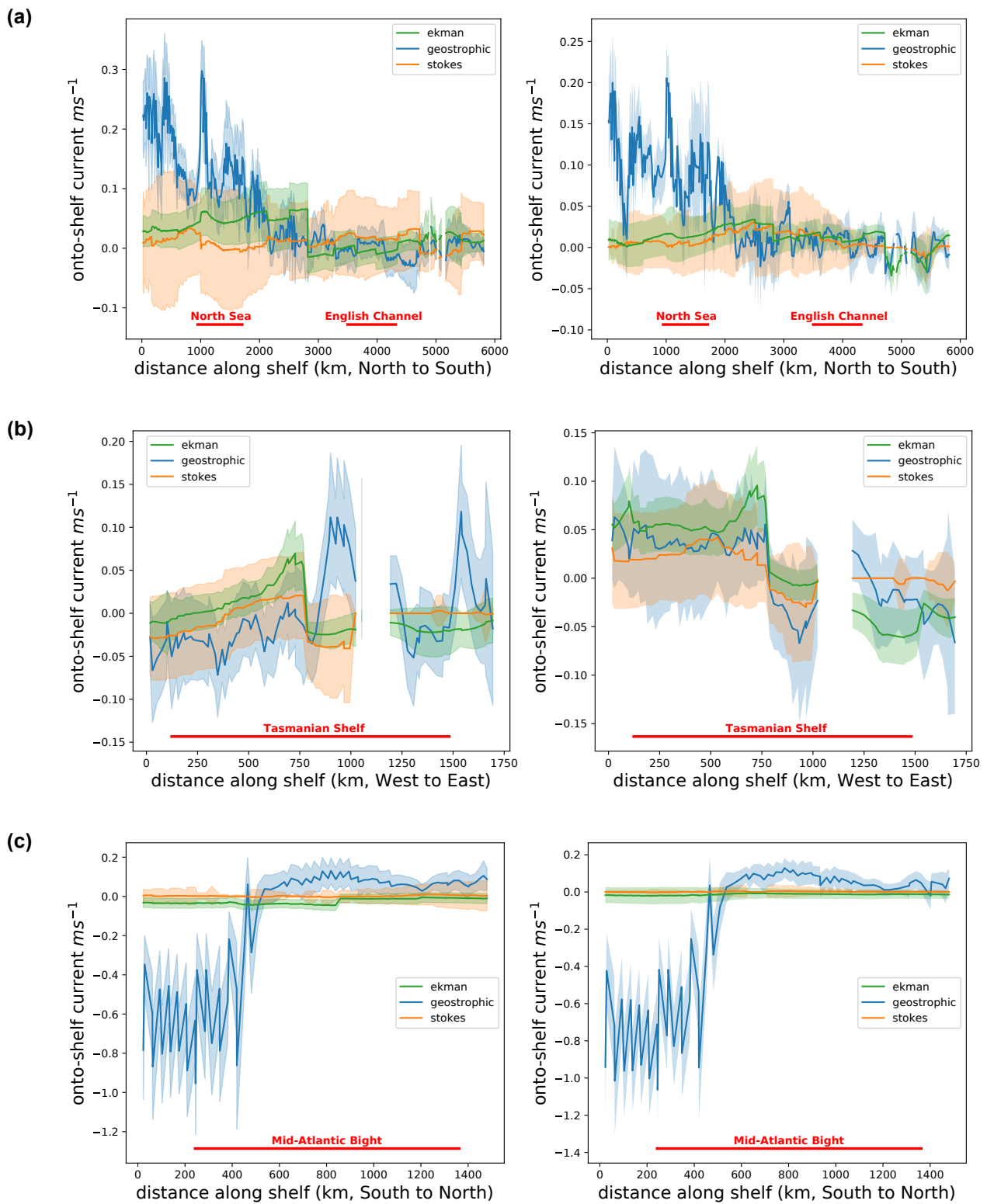

**Figure 3** Contributions to transport within the mixed layer and normal to the shelf edge due to Ekman (green) and geostrophic (blue) currents during northern hemisphere winter (January, February and March) and summer (July, August and September) for 1993 to 2016. The equivalent Stokes component normal to the shelf edge where $|\tau^{wy}| < 0.03$ N m$^{-2}$ and $H_s > 2$ m is shown in red. Shaded areas are ±1 standard deviation. Regions are a) the European shelf sea; b) the Tasmanian shelf and c) the Mid-Atlantic Bight.

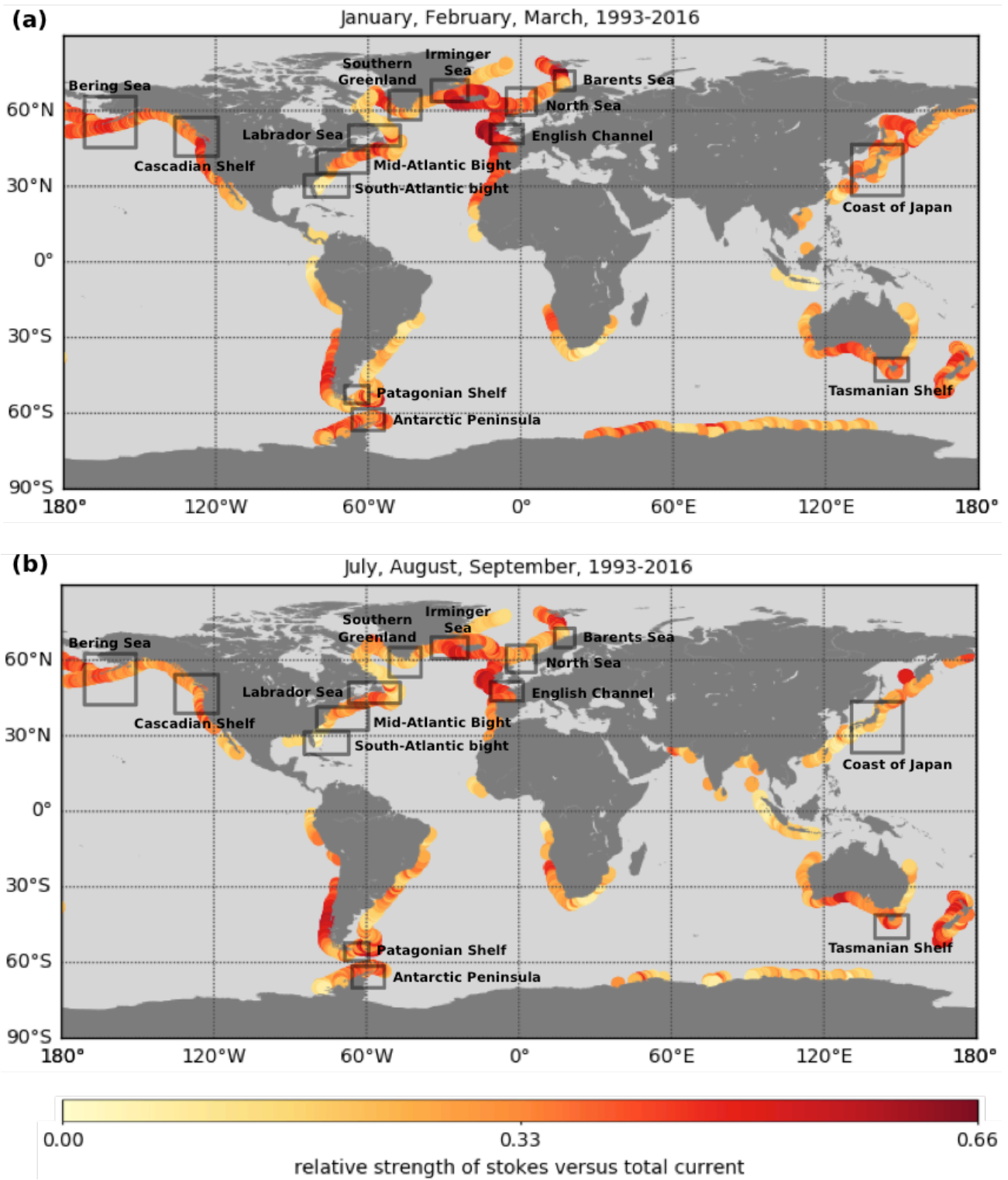

**Figure 4** Regions and times where ageostrophic Stokes drift could become important for cross-shelf transport during a) northern hemisphere winter (January, February and March) and b) northern hemisphere summer (July, August and September) for 1993 to 2016. A value of 0.0 indicates transport dominated by Ekman and/or geostrophic processes, whereas a value of 1.0 indicates transport dominated by ageostrophic Stokes. The boxes are the shelf-seas studied by Laruelle et al., (2018).

## 3.2 Case study: the role of ageostrophy in cross-shelf exchange and the potential of SKIM

Figure 5 shows the magnitude of the total onto shelf surface current for northern hemisphere winter (Figure 5a) and summer (Figure 5b) periods. The coloured bar charts i) to iv) provide the mean signed percentage of the total reference current for each boxed region that is attributable to the geostrophic, Ekman and the ageostrophic components. It is clear that the significance and strength of the ageostrophic component varies along the shelf and between seasons. For example, there is evidence that the ageostrophic component can be of a similar magnitude to, and apposing, the geostrophic component (e.g. Figure 5a, iii and iv). The horizontal black lines in the bar charts i) to iv) show the results when using the simulated SKIM data as the reference. In the majority of cases the simulated SKIM data successfully capture the different current components. There is however evidence of the SKIM data missing some of the ageostrophic component during the summer when swell waves are more likely to coincide with low wind (Figure 5b) and in especially in Figure 5a ii). These issues are likely related to the inversion of the wave Doppler within the SKIM simulation. Multiple viewing azimuths are needed to accurately calculate the wave spectrum and momentum, whereas these simulations were derived using a single azimuth. Future updates to the simulations will improve this.

## 4.0 Discussion

This study has identified that pressure driven geostrophic, wind driven Ekman and wave driven Stokes currents are all important for characterising cross-shelf transport. It also identifies when and where each component is likely to dominate, or be significant, for different shelf-seas. The results presented are consistent with previous regional studies. This analysis highlights the difficulty in using isolated *in situ* studies to characterise the wider shelf regions, as important variability that exists elsewhere along the same shelf-edge can be missed. Globally the relationship between the different current components (geostrophic, Ekman, Stokes and other ageostrophic components) and their strength can be highly variable in space and between seasons. Future studies should attempt to quantify all major components and avoid assumptions that the importance of each component is time independent.

The $|\mathbf{n}(C_E + 45°)|$ approach is used to provide an indicator of the upper bounds of $C_E$ across the mixed layer. This approach is a likely overestimate, as it does not take account of the exponential decrease of $C_E$ with depth. However the results are consistent with published studies (table 1) and there are clearly instances where Ekman processes do not dominate. Stokes drift is not necessarily aligned with the wind stress and can be supported by non-local swell systems. The Stokes drift can affect the surface mixed layer, and thus the vertical diffusion to impact the final Ekman spiral. For example, Ardhuin (2009) identified that Stokes drift can cause the deflection angle in the Ekman layer to increase into the range of 45-70°. As a first evaluation this analysis has ignored interactions between the different current components, focussing instead on identifying where each component is likely to dominate or influence. This analysis has ignored the impact of tidal flows and currents, as none of the datasets include tidal oscillations or their influences. Tidal flow influence will generally be small at the surface (Huthnance et al., 2009), but can be important for shelf exchange in some regions (Graham et al., 2018). Further work to reduce uncertainties in the definition of the mean dynamic topography at shelf-sea spatial scales would be beneficial and is possible (Rio et al., 2014).

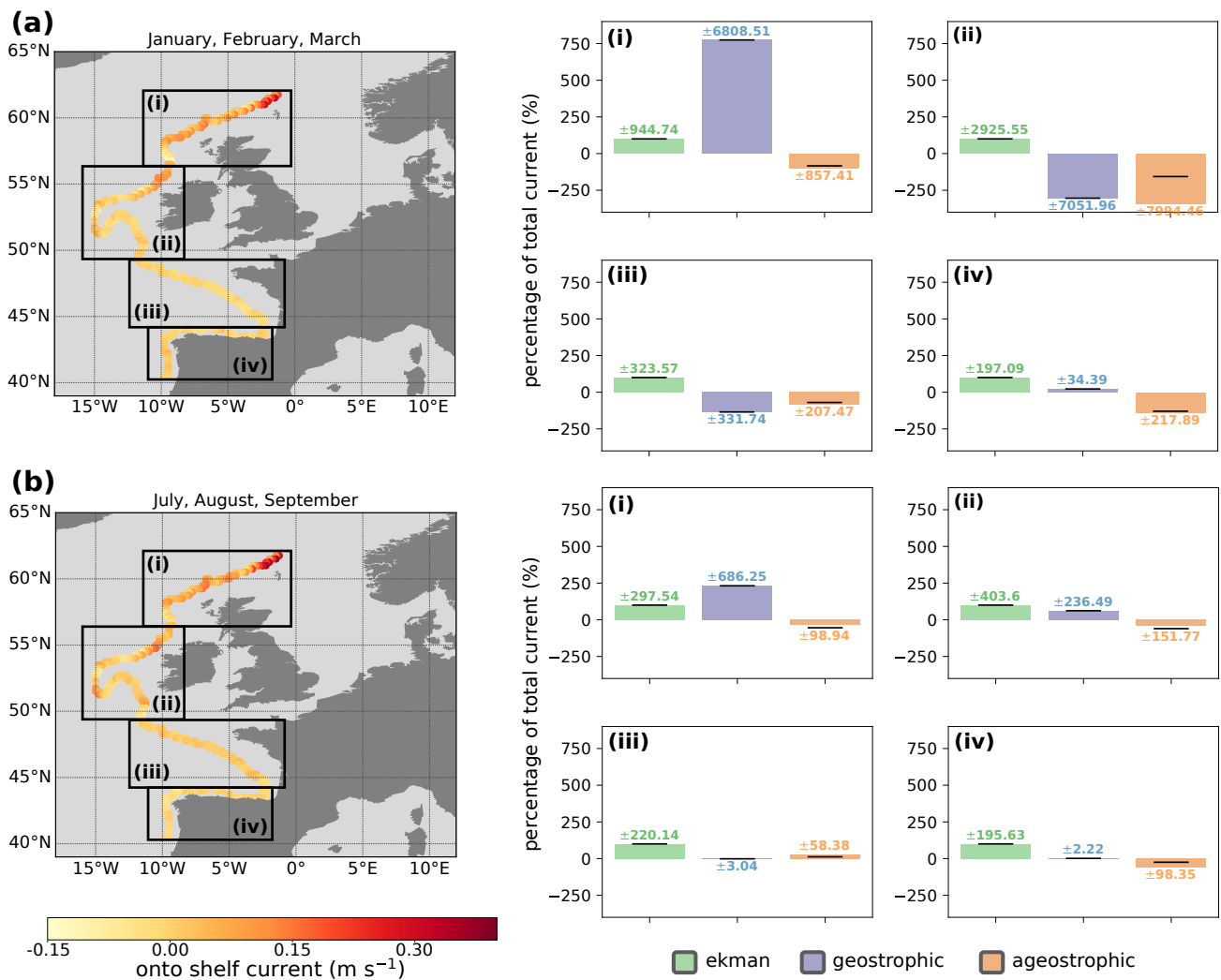

**Figure 5** Relative contributions to surface currents across sections of the European continental shelf-edge due to geostrophic (purple), Ekman (green) and ageostrophic (orange) components during a) the northern hemisphere winter (January, February and March 2012) and b) the northern hemisphere summer (July, August and September 2012) as derived from the NATL60 CJM165 simulations. The black horizontal lines show the values derived from the simulated SKIM data. Values above each bar give ±1 standard deviation.

**4.1 The implications for shelf-seas and carbon export**

Laruelle et al., (2018) identified that the rate of increase in $pCO_2$ varies for different shelf-seas. Using a large 18 year winter time *in situ* dataset they identified fifteen shelf regions where trends in $pCO_2$ were significant. Table 2 shows the $d\Delta pCO_2/dt$ rates ($\Delta pCO_2$ = air minus water $pCO_2$, t is time) and groupings from Laruelle et al., (2018) along with the corresponding winter-time mixed layer transport components, their relative strengths and the winter-time atmosphere-ocean (air-sea) $CO_2$ gas exchange (transfer) rates calculated using a wind speed gas transfer parameterisation (Nightingale et al., 2000), the GlobCurrent wind speed data and a sea surface temperature dataset (Banzon et al., 2016). Generally, those seas where a positive $d\Delta pCO_2/dt$ rate has been identified have a geostrophic dominated cross-shelf transport, so the dominant component of the cross-shelf exchange is not directly related to the dominant processes driving atmosphere-ocean exchange (i.e. wind and sea state, e.g. Blomquist et al. 2018). Thus, increased atmosphere-ocean surface exchange (and carbon accumulation) does not directly result in an increase in export at depth, implying a carbon 'bottle neck' within the shelf-sea. Those seas where the processes driving cross-shelf transport and atmosphere-ocean exchange are coupled and additive exhibit no increase in rate (i.e. $d\Delta pCO_2/dt$ = 0). In these waters any increase in atmosphere-ocean surface exchange (and carbon accumulation) can result in a corresponding increase in carbon export at depth (and hence no 'bottle neck'). There are also examples where the geostrophic flow is dominant but opposes the Ekman and Stokes components (e.g. Southern Greenland). This implies that increased surface atmosphere-ocean exchange (from wind and waves) will likely result in reduced offshore cross-shelf flow and thus increased carbon accumulation in surface waters. These results support the hypothesis that coupling between the processes driving atmosphere-ocean exchange and those driving cross-shelf transport (and hence later deep ocean export) are important for controlling carbon accumulation in the surface waters (i.e. controlling the $d\Delta pCO_2/dt$ rate). It is interesting to note that Laruelle et al., (2018) suggested that the reason for differing $d\Delta pCO_2/dt$ rates along the Mid Atlantic Bight could be due to differing hydrodynamic regimes. Figure 3c supports this conclusion, as a clear change in direction of the dominant geostrophic cross-shelf transport component is evident part way along the shelf.

It appears possible that shelf-seas exhibiting cross-shelf transport dominated by geostrophic flow (i.e. >50% geostrophic) will continue to accumulate carbon, increasing in their sink strength and ocean acidification. Whereas those shelf seas where cross-shelf transport is strongly influenced by wind and wave induced flow (and these are additive) could continue to track atmospheric values. Clearly in all cases expected future changes in wind and wave climate (e.g. Dobrynin et al., 2012; Liu, 2016) and interactions supports the need to monitor and revise off-shelf transport estimates and the size of the continental shelf-sea carbon pump.

**Table 2** Continental shelf seas groups with identified rates of change (significant decadal trends) in ΔpCO$_2$ and their controls of cross-shelf transport and strengths of atmosphere-ocean (air-sea) gas exchange. ΔpCO$_2$ rates and region names are from Laruelle et al., (2018). An increase in dΔpCO$_2$/dt implies a strengthening sink of CO$_2$ (and increasing ocean acidification), nominal or no increase implies a temporally constant sink, whereas a decrease implies a weakening sink. The Baltic Sea is not included due to no data.

| dΔpCO$_2$/dt groups from Laruelle et al., (2018) | Seas from Laruelle et al., (2018) | Winter time cross-shelf exchange μ±σ (m s$^{-1}$) | Winter time air-sea exchange μ±σ (10$^{-6}$ m s$^{-1}$) | Observations of winter control of cross-shelf transport in mixed layer | Implied conditions |
|---|---|---|---|---|---|
| High rate of increase (+2 μatm yr$^{-1}$) | North Sea (NS) Mid Atlantic Bight (MAB) Southern Greenland (SG), Antarctic Peninsula (AP) | NS 0.16 ± 0.15 MAB -0.08 ± 0.30 SG -0.13 ± 0.16 AP 0.01 ± 0.08 (weak to medium exchange) | NS 22.47 ± 17.94 MAB 13.08 ± 8.64 SG 18.39 ± 17.17 AP 13.33 ± 12.36 (medium to high air-sea exchange). | The dominant cross-shelf flow is geostrophic and therefore independent of the dominant processes driving air-sea exchange (For NS, MAB, SG, AP ≥53% geostrophic). Offshore geostrophic surface flow opposes Ekman and Stokes components (SG) and so increases in processes driving air-sea exchange imply reduced cross-shelf flow. | Imbalance between cross-shelf exchange and air-sea exchange (bottle neck in offshore transport). |
| Moderate rate of increase (e.g. +0.5 to 1.0 μatm yr$^{-1}$) | Irminger Sea (IS), Labrador Sea (LS), Coast of Japan (CoJ), Cascadian Shelf (CS), South Atlantic Bight (SAB). | IS: -0.04 ± 0.09 LS 0.01 ± 0.18 CoJ 0.003 ± 0.17 CS -0.06 ± 0.14 (weak exchange) SAB -0.50 ± 0.56 (very high exchange) | IS: 17.19 ± 17.17 LS 21.89 ± 18.69 CoJ 18.72 ± 12.39 CS 12.19 ± 12.03 (medium to high air-sea exchange). SAB 5.47 ± 4.78 (low air-sea exchange) | The dominant cross-shelf flow is geostrophic, and therefore independent of the dominant processes driving air-sea exchange (LS, CoJ ≥ 54% geostrophic). Surface flow is offshore with high air-sea exchange (IS, CS) implying that a portion of the increased surface water carbon from high air-sea exchange is retained as no deep-water export. Very high offshore surface exchange combined with low air-sea exchange (SAB). No deep-water export. | Imbalance between cross-shelf exchange and air-sea exchange (bottle neck in offshore transport). |
| Nominal or no increase (in water pCO$_2$ tracks atmosphere pCO$_2$) | English Channel (EC), Barents Sea (BaS), Tasmanian Shelf (TS) | EC 0.01 ± 0.08 BaS 0.23 ± 0.13 TS 0.04 ± 0.12 (weak to high exchange) | EC 18.97 ± 19.00 BaS 12.14 ± 10.64 TS 22.53 ± 14.75 (medium to high air-sea exchange). | Equal dominance and additive geostrophic and Ekman cross-shelf flow (EC, TS), or high cross-shelf flow is geostrophic dominated surface flow and additive with Ekman and Stokes (BaS, 71%, 11%, 19%). | Cross-shelf exchange is balanced by air-sea exchange (no bottle neck). |
| Moderate decrease (-0.2 to -1.1 μatm yr$^{-1}$) | Patagonian shelf (PS), Bering Sea (BeS) | PS -0.17 ± 0.18 (high exchange) BeS -0.003 ± 0.11 (very weak exchange) | PS 33.73 ± 18.81 (very high air-sea exchange). BeS 10.19 ± 8.14 (low to medium air-sea exchange). | Surface flow dominated by consistently strong off-shelf geostrophic component implying that increased surface water carbon from elevated air-sea exchange is immediately exported away from shelf waters (PS). Very weak, but variable offshore surface flow. Geostrophic, Ekman, Stokes components are additive (BeS, 44%, 23%, 33%). | Much faster cross-shelf exchange compared with air-sea exchange (high deep water export), or surface flow is offshore (no deep water accumulation or export). |

## 4.2 The potential for routine monitoring of the continental shelf sea pump

Most ocean hydrodynamic and ecosystem models assimilate ocean observations of temperature, salinity and sea level allowing them to identify global ocean circulation based on geostrophy. However, theory, observations and the work presented here demonstrate that the circulation is not in geostrophic balance along shelf-edges (Niiler, 2009). Consequently, each model will produce different ageostrophic flows and exchanges within these areas, dependent upon the chosen model vertical structure, the underlying bathymetry dataset and turbulence parameterisation (Niiler, 2009). Furthermore, recent work has highlighted that key processes in cross-shelf exchange only begin to be resolved at spatial scales of the order ~1 km, so coarse scale global models used to assess carbon cycles are unable to capture this exchange (Graham et al., 2018). Thus, observations are needed to constrain and challenge model choices and parameterisations if we are to be able to monitor and predict export from shelf-seas. The results presented here suggest that SKIM would be able to provide surface observations of the geostrophic and ageostrophic surface velocities, and their interactions, suitable for parameterising and challenging such models.

## 5.0 Conclusions

A general lack of appropriate observations means that the combined influence of wind, currents and waves in modulating cross-shelf transport is not well studied. Here we have taken a pragmatic approach to identify the relative importance of the different ocean current components in influencing global cross-shelf surface transport, which in turn, through the need for mass balance, controls the export of water and thus carbon at depth. The accumulation of $CO_2$ in the surface waters in global shelf-seas appears to be variable and increasing. This work supports the hypothesis that imbalances and differences in the processes driving atmosphere-ocean exchange and carbon export at depth are controlling the change in shelf-sea $CO_2$ sinks and their acidification; therefore both types of exchange warrant monitoring. This monitoring requires a synergy approach between measurements and hydrodynamic modelling. The proposed Sea Surface Kinematics Multiscale monitoring satellite mission (SKIM) appears capable of providing the measurements essential for parameterising, constraining and challenging such a monitoring approach.

**Acknowledgements**

The authors acknowledge funding from the European Space Agency (ESA) through the Sea surface KInematics Multiscale monitoring (SKIM) Mission Science study (contract 4000124734/18/NL/CT/gp) and the ESA SKIM Scientific Performance Evaluation study (contract 4000124521/18/NL/CT/gp). Bathymetry data are from GEBCO_2014 grid, version 20150318 and can be downloaded from: www.gebco.net.

**Supplemental material**

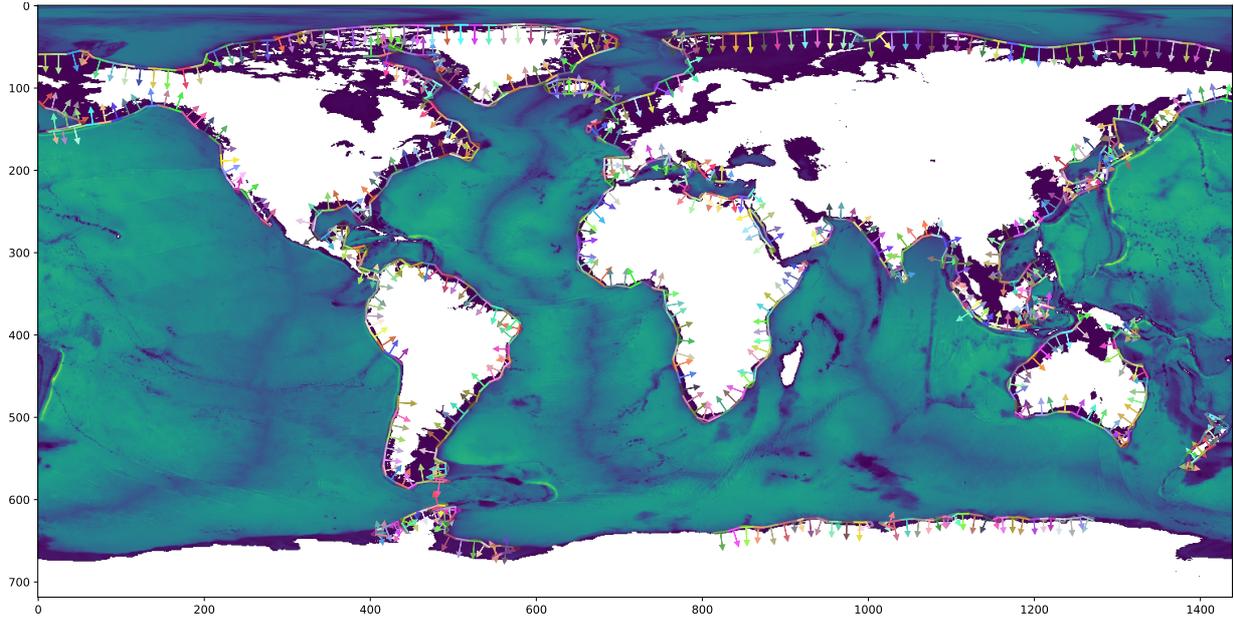

**Figure S1:** Global shelf straight line approximations and onto-shelf normal vector directions.

**Table S1** Sensitivity of calculated cross-shelf transport results to choice of shelf boundary depth. Main results using the 500 m shelf boundary depth (and 600 m as the deep contour) are compared to the mean generated across 9 scenarios using shelf boundary depths of 300, 350, 400, 450, 500, 550, 600, 650 and 700 metres. For the purpose of calculating onto-shelf direction the deep contour depth was taken as the shelf boundary depth plus 100 m. Standard deviations are calculated and shown for each scenario.

| Region (Reference) | *In situ* study period (*In situ* across-shelf current mean or range, m s$^{-1}$) | Calculated across-shelf current (m s$^{-1}$) using just 500 m depth | Components (proportion within the mixed layer) using just 500 m depth | Calculated cross-shelf current (m s$^{-1}$) using all depths | Components (proportion within the mixed layer) using all depths |
|---|---|---|---|---|---|
| Scottish Hebrides, European shelf (Painter et al., 2016) | September 2014 (0.15 – 0.36) | 0.19 (+/- 0.05) | Geostrophic (0.21 +/- 0.15) Ekman (0.16 +/- 0.03) Stokes (0.62 +/- 0.15) | 0.19 (+/- <0.01) | Geostrophic (0.21 +/- 0.01) Ekman (0.16 +/- <0.01) Stokes (0.62 +/- 0.01) |
| South Atlantic Bight (Yuan et al., 2017) | January to March, 2002 to 2014 | -0.53 (+/- 0.55) | Geostrophic (0.9 +/- 0.14) Ekman (0.10 +/- 0.14) Stokes (0.01 +/- 0.04) | -0.40 (+/- 0.09) | Geostrophic (0.86 +/- 0.04) Ekman (0.13 +/- 0.04) Stokes (0.01 +/- <0.01) |
| South Atlantic Bight (Yuan et al., 2017) | July to September, 2002 to 2014 | -0.58 (+/- 0.61) | Geostrophic (0.92 +/- 0.11) Ekman (0.08 +/- 0.11) Stokes (<0.01 +/- 0.01) | -0.42 (+/- 0.11) | Geostrophic (0.88 +/- 0.03) Ekman (0.12 +/- 0.03) Stokes (<0.01 +/- <0.01) |
| Mid Atlantic Bight (Fewings et al., 2008) | All season mean, June 2001 to May 2007 (0.01) | 0.05 (+/- 0.09) | Geostrophic (0.63 +/- 0.24) Ekman (0.28 +/- 0.23) Stokes (0.09 +/- 0.17) | 0.05 (+/- <0.01) | Geostrophic (0.62 +/- 0.01) Ekman (0.29 +/- <0.01) Stokes (0.09 +/- <0.01) |
| Mid Atlantic Bight (Fewings et al., 2008) | October-March, 2001 to 2007 (0.01) | 0.07 (+/- 0.09) | Geostrophic (0.59 +/- 0.24) Ekman (0.26 +/- 0.21) Stokes (0.15 +/- 0.20) | 0.07 (+/- <0.01) | Geostrophic (0.58 +/- 0.01) Ekman (0.26 +/- <0.01) Stokes (0.16 +/- <0.01) |
| Mid Atlantic Bight (Fewings et al., 2008) | April-September, 2001 to 2007 (<0.06) | 0.04 (+/- 0.08) | Geostrophic (0.66 +/- 0.23) Ekman (0.30 +/- 0.23) Stokes (0.05 +/- 0.13) | 0.04 (+/- <0.01) | Geostrophic (0.65 +/- 0.01) Ekman (0.30 +/- 0.01) Stokes (0.05 +/- <0.01) |
| California coast (Woodson, 2013) | April-September, upwelling season, 2004 to 2009 | 0.13 (+/- 0.06) | Geostrophic (0.20 +/- 0.14) Ekman (0.74 +/- 0.20) Stokes (0.06 +/- 0.14) | 0.13 (+/- 0.01) | Geostrophic (0.20 +/- 0.01) Ekman (0.74 +/- 0.01) Stokes (0.06 +/- <0.01) |
| California coast (Woodson, 2013) | October-March, non-upwelling season, 2004 to 2009 | 0.08 (+/- 0.07) | Geostrophic (0.25 +/- 0.16) Ekman (0.39 +/- 0.21) Stokes (0.36 +/- 0.19) | 0.08 (+/- <0.01) | Geostrophic (0.24 +/- 0.01) Ekman (0.40 +/- 0.02) Stokes (0.36 +/- 0.01) |
| Eastern Indian Ocean, Western Australia (Waite et al., 2016) | May 2006 (spring) | 0.11 (+/- 0.21) | Geostrophic (0.8 +/- 0.18) Ekman (0.09 +/- 0.08) Stokes (0.11 +/- 0.11) | 0.10 (+/- 0.03) | Geostrophic (0.76 +/- 0.04) Ekman (0.11 +/- 0.02) Stokes (0.14 +/- 0.03) |
| East China Sea (Wei, 2013) | April 1987 to January -2010 (calculated for January 1993 to January 2010) | -0.14 (+/- 0.26) | Geostrophic (0.78 +/- 0.22) Ekman (0.22 +/- 0.22) Stokes (<0.01 +/- 0.03) | -0.10 (+/- 0.09) | Geostrophic (0.73 +/- 0.07) Ekman (0.26 +/- 0.08) Stokes (<0.01 +/- <0.01) |